%% file: versao_final.tex
\documentclass[review, 3p]{elsarticle}
\usepackage[utf8]{inputenc}

\usepackage{lineno,hyperref}

\usepackage{amssymb, amsmath}

\usepackage{numcompress}
\usepackage{lineno}

\usepackage{float}
\usepackage{url}

  \usepackage[caption=false,font=normalsize,labelfont=sf,textfont=sf]{subfig}

\usepackage[figuresright]{rotating}

\usepackage{listings}
\usepackage{xcolor}
\usepackage{color}

\definecolor{green}{rgb}{0,0.5,0}
\definecolor{blue}{rgb}{0,0.5}
\definecolor{yellow}{rgb}{0.5,0.5,0}
\definecolor{gray}{rgb}{0.5,0.5,0.5}

\modulolinenumbers[5]

\begin{document}

\begin{frontmatter}

\title{A Scalable Shared-Memory Parallel Simplex for Large-Scale Linear Programming}





\address[ifrn]{Federal Institute of Education, Science and Technology of Rio Grande do Norte, BR 405, 59900-000, Pau dos Ferros, Brazil}
\address[ufrnadd]{Universidade Federal do Rio Grande do Norte, 59078-970, Natal, Brazil}
\address[montreal]{Computer and Software Engineering Department, École Polytechnique de Montréal, Montréal, Canada}

\author[ifrn,ifrnaddress]{Demetrios~A.~M.~Coutinho\corref{demetrios}\fnref{referencia}}
\fntext[referencia]{Corresponding author.}
\ead{demetrios.coutinho@ifrn.edu.br}

\author[ufrnadd]{Felipe~O.~Lins~e~Silva}
\ead{lins.flp@gmail.com}

\author[montreal]{Daniel~Aloise}
\ead{daniel.aloise@polymtl.ca}

\author[ufrnadd]{Samuel~Xavier-de-Souza}
\ead{samuel@dca.ufrn.br}

\begin{abstract}
The Simplex tableau has been broadly used and investigated in the industry and academia. With the advent of the big data era, ever larger problems are posed to be solved in ever larger machines whose architecture type did not exist in the conception of this algorithm. In this paper, we present a shared-memory parallel implementation of the Simplex tableau algorithm for dense large-scale Linear Programming (LP) problems for use in modern multi-core architectures. We present the general scheme and explain the strategies taken to parallelize each step of the standard simplex algorithm, emphasizing the solutions found to solve performance bottlenecks. We analyzed the speedup and the parallel efficiency for the proposed implementation relative to the standard Simplex algorithm using a shared-memory system with 64 processing cores. The experiments were performed for several different problems, with up to 8192 variables and constraints, in their primal and dual formulations. The results show that the performance is mostly much better when we use the formulation with more variables than inequality constraints. Also, they show that the parallelization strategies applied to avoid bottlenecks lead the implementation to scale well with the problem size and the core count up to a certain limit of problem size. Further analysis showed that this scaling limit was an effect of resource limitation. Even though, our implementation was able to reach speedups in the order of 19$\times$.
\end{abstract}


\end{frontmatter}



\input{tex/introduction}

\input{tex/simplex.tex}
\input{tex/parallel.tex}
\input{tex/results.tex}
\input{tex/consideration.tex}
\bibliographystyle{model6-num-names}
\bibliography{References}

\textbf{Demetrios~A.~M.~Coutinho} holds a bachelor's degree in Computer Engineering, with the emphasis in Industrial Automation from the Universidade Federal do Rio Grande do Norte (UFRN) (2011) and a master's degree in Electrical Engineering and Computation in the area of Linear Programming and Parallel Computing from UFRN (2014).

\textbf{Felipe O. Lins e Silva} has a Bachelor of Science Degree in Science and Technology awarded at Universidade Federal do Rio Grande do Norte (UFRN) in 2018. Currently Currently, he is a B.Sc. Student in Computer Engineering and his interest are in Data Science, Signal Processing and Quantitative Finances.

\textbf{Dr Daniel~Aloise} is  assistant professor at the Computer and Software Engineering Department at the École Polytechnique de Montréal. He obtained his PhD in Applied Maths from Polytechnique Montréal in 2009. He is member of the GERAD (Group for Research in Decision Analysis) and fellow of the  Canada Excellence Research Chair in Data Science for Real-Time Decision-Making. His research interests include data mining, optimization, mathematical programming and how these disciplines interact to tackle problems in the Big Data era. 

\textbf{Samuel~Xavier-de-Souza} is associate professor at Universidade Federal do Rio Grande do Norte-UFRN, and director of the High-Performance Computing Center-NPAD. He became doctor in electrical engineering from Katholieke Universiteit Leuven in 2007. In 2016, he became a Royal Society-Newton Advanced Fellow for his work on Energy-Efficient Software. His interests are on parallel computing, energy-efficient software, and related applications.

\end{document}

%% file: tex/introduction.tex
\section{Introduction}

Parallel processing has arisen as an important leverage for enabling high-performance scientific computing and sustained productivity. Since mid 2000's, we have been facing a paradigm change where the computers are not being produced with only one processing core. This trend has being called the Multi-core Era~\cite{koch, borkar}, which encompasses the principle of doubling the number of processing cores in a single chip with each new technology generation. In this context, several existing algorithms needs to be revisited in order to become ready for this era and, although many have already gone to this process, many other remain to be analyzed and optimized.

One of the challenges to make algorithms ready for the multi-core era is the number of new architectures to target for. While in the single-core world there was a consolidated architecture to target, in the current era there are many different options that present different performance for the same parallel algorithm and that respond to architectural-based optimizations very specifically.

Another challenge involves the size of the problems that need to be solved today. In the era of Big Data, employing traditional algorithms to solve problems many orders do magnitude larger than they were designed for may lead to inefficient solutions, in terms of economic cost, or impractical solutions, in terms of time.

In this context, this work aims to improve the parallel performance of Linear Programming (LP) solvers for large dense problems on multi-core systems. 
The simplex method is widely used academically and commercially to solve LP problems. 

There has been many studies with parallel versions of the simplex method. Many of those have focused on the revised simplex method due to the advantage of solving sparse problems and being most effective where the number of variables is much larger than the amount of restrictions \cite{implmnt}. However, the revised method is not well suited for parallelization \cite{art1061}. Eckstein et. al \cite{cm2} describe three simplex parallel implementations, including a revised method for dense LP problems on the Connection Machine CM-2. They showed that their implementation can yield faster execution times on the CM-2 than on a workstation from the same era and processing power. Thomadakis and Liu \cite{steepest} worked on parallelizing the Standard and Dual simplex algorithm, comparing them on two versions of Maspar, the MP-1 and the MP-2. Their results show that as the problems size increases the speedup obtained by 1,024 processors (MP-1) and 16,384 processors (MP-2) is about 100$\times$ and about 1000$\times$, respectively, over the sequential standard simplex algorithm. Hall and McKinnon \cite{ms95050} studied an asynchronous variant of the revised simplex method on the Cray T3D machine. They present different ways for improving the performance of this algorithm, using distinct T3D routines. They also emphasize the potential of this variant for shared memory processors, instead of the parallel distributed implementation on the Cray T3D.




%
%
%
%

In 2000, Maros and Mitra \cite{ssx} presented a cooperative parallel version of the sparse simplex method for linear programs. They have adopted a distributed memory multiprocessor, the Parsytec CC12 computer with 12 processors and 64 Mbyte RAM per processor node. The objective was to reserve and dedicate each processor node to execute one of the chosen column selection strategies. They identified that a considerable improvement in efficiency can be achieved by the changing strategy depending on the problem behavior.

Later, in 2009, Yarmish and Slyke \cite{artdistsimplex} presented a scalable simplex implementation for large-scale linear programming problems using 7 workstations connected by Ethernet. Their parallel standard algorithm showed to be more efficient than the revised method. In the same year, Ploskas et al. \cite{ploskas} presented a parallel implementation of the standard simplex algorithm using a personal computer with two cores. Due to dense matrices and heavy communication, the ratio of computation to communication was extremely low. Their computational results show that a linear speedup is hard to achieve even with carefully selected partitioning patterns and communication optimization.

In 2013, a parallelization of the revised simplex method for large extensive forms of two-stage stochastic linear programming problems was proposed by Lubun et al.~\cite{stochastic}.  They have developed the linear algebra techniques necessary to exploit the dual block-angular structure of an LP problem inside the revised simplex method and a technique for applying product form updates efficiently in a parallel context. Their result show that the implementation was most effective on large instances that could be considered very difficult to solve using the revised simplex algorithm. In the same year, a parallel implementation of the revised simplex algorithm using OpenMP in a shared memory multiprocessor architecture was presented by Ploskas~\cite{openmpRevised}. It focused on the reduction of the time taken to perform the basic matrix inverse. Their computational results with a set benchmark problems from Netlib\footnote{The Netlib repository contains freely available softwares, documents and databases of interest to the scientific computing, numerical and other communities. The repository is maintained by AT \& T Bell Laboratories, from the University of Tennessee and Oak Ridge National Laboratory. For more information: \url{http://www.netlib.org/}} reported a 1.79 average speedup on a machine with 4 cores. 

A few yeares later, in 2015, Ploskas and Samaras \cite{gpu} proposed two efficient GPU-based implementations of the revised method and a primal-dual exterior point simplex algorithm. Both parallel algorithms were implemented in MATLAB using MATLAB's Parallel Computing Toolbox. They performed a computational study on large-scale randomly generated optimal sparse and dense LPs and found that both GPU-based algorithms outperformed MATLAB's interior point method. The GPU-based primal-dual exterior point simplex algorithm shows an average speedup of 2.3 over MATLAB's interior point method on a set of benchmark LP problems.


Remarkably, the major efforts to parallelize simplex-based solvers for LP problems focused on the revised method while strategies involving the standard method are scarcer. This is probably due to the fact that the revised method is more efficient for sparse LP problems~\cite{implmnt,artdistsimplex}. 
Nonetheless, dense LP problems do occur in a number of applications~\cite{cm2}. Hence, since the revised method is less suitable for parallelization~\cite{proc-519254,art1061}, 
and being parallelism the major means of performance increase nowadays, efforts toward the development of scalable versions of the standard simplex method are opportune.

In general, efforts to parallelize conventional algorithms like the simplex and many others~\cite{Cuomo2017,Licht2018} are a necessary nowadays in order to consider the limitations brought by both multi-core and big data eras and devise the appropriate optimizations and specializations to the target architecture. This effort also involves figuring out the best configuration for execution depending on the problem at hand.



In this paper, we present a scalable parallel implementation of the standard simplex algorithm, based in~\cite{Coutinho2013}, for large-scale dense LP problems using OpenMP. We detail how we parallelize each step of the algorithm detailing the strategies used to avoid performance bottlenecks that resulted in better efficiency when compared to the previous version~\cite{Coutinho2013}. We analyzed the performance gains of our parallel implementation over the sequential version, using a shared memory computer with 64 processing cores. We measured the speedups for different numbers of variables and constraints for several different problems. With that, we found out for which cases it may be more advantageous to solve problems with more variables than constraints, or vice versa, which implies in a better choice for using the primal or the dual formulation of the LP problem. The performance results show up to 19$\times$ factor of improvement for the proposed parallel implementation over the sequential standard simplex. The whole project is publicly available at access~\url{https://gitlab.com/lappsufrn/MulticoreParallelSimplex}.



This paper is organized in the following way. In Section 2, the Standard Simplex Algorithm is described. In Section 3, we present the proposed parallel scheme and its implementation details. In Section 4, we show the scalability and performance results. And, finally, in the Conclusion, we make our final considerations about our main contributions.

%% file: tex/simplex.tex
\section{The Standard Simplex Algorithm}
\label{simplex}
Several methods are available for solving Linear Programming, among which the Simplex algorithm is the most widely used \cite{gpu}. Solving an LP problem using the Simplex method requires it to be in the standard form.

    \begin{center}
    \textbf{Standard Form}    
    \end{center}
	\begin{equation}
		\begin{aligned}
		& \underset{x}{\text{maximize}}
		& & z = \mathbf{c}^\intercal \mathbf{x},\\
		& \text{subject to}
		& &\mathbf{Ax} = \mathbf{b},\mathbf{b} \geq 0\\
		&&&\mathbf{x} \geq 0, 
		\end{aligned} 		
		\label{padrao}
	\end{equation}
where $\mathbf{A}=[a_{ij}] \in \mathbb{R}^{m \times n}$, $\mathbf{b} \in \mathbb{R}^{m}$ and $\mathbf{c} \in \mathbb{R}^n$ are known coefficients. The simplex algorithm optimizes (\ref{padrao}) over the decision variables $\mathbf{x} \in \mathbb{R}^{n} $.


The canonical form of a linear maximization problem assumes inequality constraints of the
type $\leq$. As such, their standardization requires the addition of slack variables introduced to transform inequality constraints into equality constraints. Our simulated experiments consider linear maximization problems in canonical form, i.e., with $m$ inequality constraints of type $\leq$ . Therefore, their standardization naturally provides an initial basis for the simplex method in which the basic variables correspond to the added slack variables.

The following steps describes the simplex method.

\begin{enumerate}[Step 1.]
	\item \textbf{Build the initial tableau:} the Simplex data is placed in a tableau, which contains a row for the objective function--written as $z - \mathbf{c}^\intercal = 0$. The original variables have indexes in the range $[1,n]$, while the columns of the slack variables have indexes in the range $[n+1,n+m]$. 
	Table~\ref{tabquadro} presents the structured initial tableau.
	\begin{table}[!ht]
		\renewcommand{\arraystretch}{1.3}
		\caption{Initial tableau for the standard Simplex}
		\label{tabquadro}
		\centering
			\begin{tabular}{  c  c  c  c  c  c  c  c  c  c  }
				x$_{1}$ & x$_{2}$ & $\cdots$ & x$_{n}$ & x$_{n+1}$ & x$_{n+2}$ & $\cdots$ & x$_{n + m}$ & RHS\\
				\hline				
				$a_{11}$ & $a_{12}$ & $\cdots$ & $a_{1n}$ & 1 & 0 & $\cdots$ & 0 & $b_{1}$ \\
				$a_{21}$ & $a_{22}$ & $\cdots$ & $a_{2n}$ & 0 & 1 &  $\cdots$ & 0 & $b_{2}$ \\
				$\vdots$ & $\vdots$ & $\vdots$ & $\vdots$ & $\vdots$ & $\vdots$ & $\vdots$ & $\vdots$ & $\vdots$ \\
				$a_{m1}$ & $a_{m2}$ & $\cdots$ & $a_{mn}$ & 0 & 0 & $\cdots$ & 1 & $b_{m}$ \\
				$-c_{1}$ & $-c_{2}$ & $\cdots$ & $-c_{n}$ & 0 & 0 & $\cdots$ & 0 & 0 \\				
				\hline            
			\end{tabular}
	\end{table}
	
	\item \textbf{Determine a nonbasic variable to become basic:} If all reduced costs $-c_{j} \geq 0, \forall j \in [1,n]$,
	the optimal solution was found and the simplex method stops. Otherwise, choose a pivot column $k$ such that $\overline{c}_{k}$ is the maximum absolute value from all $\overline{c}_{j}  < 0$.
	
	\item \textbf{Find a basic variable to leave the current basic solution:} For each positive element $a_{ik}$, $\forall i = [1,m]$,
	calculate the ratio $\rho_{i} = \frac{b_{i}}{a_{ik}}$. Choose a pivot row $l$ that corresponds to the smallest $\rho_{i}$. If $a_{ik} \leq 0, \forall i = [1,m]$,  then halt; the problem is unbounded.
%
%
    \item \textbf{Pivot:} Divide the pivot row by $a_{lk}$, where $a_{lk}$ is the pivot coefficient found by the previous steps. Thus, the pivot element will be equal to one.
    
    \item \textbf{Do the elementar operation:} $row_{i} = row_{i} \;-\; a_{ik} row_{l}, \forall i \in [1,\cdots,l-1,l+1,\cdots,m+1]$, where $row_{l}$  is the pivot row. This step is important to transform all elements of the pivot column into zeros, except for the pivot element. This operation is applied to all rows $i \neq l$, including the objective function row.
		
%
\end{enumerate}

%% file: tex/parallel.tex
\section{Parallel Scheme and Implementation}

\begin{figure*}[!ht]
	\centering	
	\includegraphics[width=\textwidth]{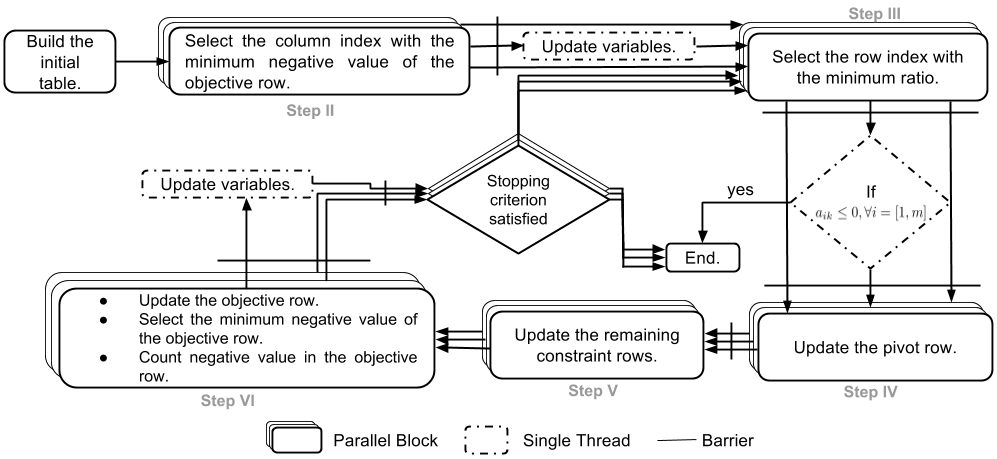}
	\caption{Flowchart of the proposed Simplex algorithm.}
	\label{fluxograma}
\end{figure*}

The proposed parallel scheme is fundamentally related to the standard simplex scheme presented in Section~\ref{simplex}. A few modifications to the original scheme was necessary to ensure the code optimizations that aimed to solve performance bottlenecks such as synchronization and load balancing. The resulting scheme contains 6 steps instead of 4. Most of the steps in the proposed parallel scheme retained the same objectives of the original steps. Mainly, steps II and IV had significant changes in their objectives, which are jointly covered in the new scheme by steps II, IV, V and VI. \figurename~\ref{fluxograma} presents an illustration of the proposed parallel scheme containing its control flow. The parallel scheme was implemented in C++ using the OpenMP~\cite{openMP} programming model. Algorithm~\ref{code1} presents the code skeleton of the scheme containing the main OpenMP parallelization directives used in each step of the algorithm. 

Step I of the scheme assembles the initial simplex table with all constraints and variables. Then, the threads are created and remain active until the end of the execution in order to avoid extra overhead of creating and destroying threads. 

\figurename~\ref{fluxograma} and Algorithm~\ref{code1} show how step II was modified to enable less synchronization. The step II runs before the main loop structure (lines 3 to 6). Each thread works locally on a set of columns to find the index of the column holding the maximum absolute value among the negative elements in the objective row. Then, it is reduced to a global variable by using the {\tt reduction(maximum:max)} clause (line 3). The {\tt schedule(guided,chunk)} clause (line 3) is used to assign the iterations dynamically to the team of threads in chunks that decrease in size as the iteration number progresses. Each thread executes a chunk of iterations, then requests another chunk, until no chunks remain to be assigned. The implicit barrier of the {\tt for} directive (line 3) is necessary to ensure that this step completes before the global column index is used in the next step. Using the {\tt single} directive modified by the  {\tt nowait} clause (line 7), a single thread updates control data before asynchronously joining the other threads in the next step.




Step III has a similar structure as step II, except for the threads working with rows instead of columns. The loop selects the row with the minimum ratio, as explained in section~\ref{simplex}, and then it is globally reduced by the {\tt reduction(minimum:min)} clause (line 10). An implicit barrier synchronizes the global row among the threads (line 10). Another barrier before the step IV and after the {\tt single} directive (line 14), is necessary to ensure access to the correct pivot row and column for all threads. The number of negative values in pivot column are counted locally and the reduced to a global count by using the  {\tt reduction(+:count)} clause (line 10). Using the {\tt single} directive which contains an implicit barrier, a single thread verify if the variable count has the same amount of constraints - condition of the unbounded solution.

Step IV updates the pivot row in the same way as the original scheme described in Section~\ref{simplex}. The implicit barrier at the end of the {\tt for} directive (line 16) ensures that this step completes before the values in this row are used in the next step. 

Step V updates the remaining constraint rows, as described in section \ref{simplex}, in a loop without the {\tt for} directive implicit barrier by using the {\tt nowait} clause (line 20). A barrier is not necessary because the next step is independent of this one. , Since there are no loop-carried dependencies which would prevent consecutive iterations of the following loop from executing concurrently with SIMD (single instruction multiple data) instructions, the {\tt \#pragma GCC ivdep} (line 23) is introduced to allow the loop to be vectorized.

Step VI is a new step focusing in a better performance and synchronization. It has the same structure of step II. However, It performs three operations in the same loop. First it updates the objective row, then it selects the maximum absolute value among the negative elements in the objective row, and then it counts the negatives values of the objective row. A barrier ensures that the counting is correct before the stopping criterion is assessed. If there are negative values in the objective row, the stopping criterion is not reached and the main loop restarts in the step III in order to select the row with the minimum ratio. Before restarting the loop, there is a {\tt single} directive (line 32) with a implicit barrier to ensure that the global control variables are reset to prevent steps III and VI to use outdated values. The number of negative values are counted locally and then reduced to a global count by using the {\tt reduction(+:count)} clause (line 26).  

\begin{lstlisting}[label=code1,columns=fullflexible,caption=OpenMP code skeleton of the parallel scheme.]
#pragma omp parallel <data scope definitions>
{
    #pragma omp for schedule(guided,chunk) reduction(maximum:max) // Step II starts here. 
    for (j = 0; j <= column_size; j++) { 
        /* Select the column with the minimun negative value in the objective row. */
    }
    #pragma omp single nowait
    /* Update Variables. */
    do {
        #pragma omp for reduction(+:count) reduction(minimum:min)   schedule(guided,chunk)// Step III starts here.
        for (i = 0; i < row_size; i++) {
            /*Select the row with the minimum ratio.*/
        }
        #pragma omp single
        /* verify condition of unbounded solution */
        #pragma omp for // Step IV starts here.
        for (j = 0; j <= column_size; j++) {
            /* Update the pivot row. */
        }
        #pragma omp for nowait // Step V starts here.
        for (i = 0; i < row_size; i++) {
          /* Update the remaining constraint rows. */
          #pragma GCC ivdep
          for (j = 0; j <= column_size; j++)           
        }
        #pragma omp for reduction(+:count) reduction(maximum:max) schedule(guided,chunk) // Step VI starts here.
        for (j = 0; j <= column_size; j++) {
            /* Update the objective row */
            /* Select the maximum  absolute value among the negative elements in the objective row.*/
            /* Count negative values in the objective row (count). */
        }
        #pragma omp single 
        /* Update Variables. */
    } while(count > 0)
}
\end{lstlisting}


%% file: tex/results.tex
 \section{Experimental Results and Discussion}

In this section we present some numerical results on various randomly generated problems. The problems are generated using the following Octave\footnote{GNU Octave is a high-level language, primarily intended for numerical computations.} code\cite{Ketabchi2012}:

\begin{lstlisting}[language=Octave,label=code2,columns=fullflexible,caption= Generates random solvable LP Problem]
function [lpp,f,A,b] = gen_rand_deme(m,n,d)
    c = zeros(n,1);
    A =[];
    b = [];
    while c >=0
        pl=inline('(abs(x)+x)/2');
        A=sprand(m,n,d);
        A=A*50;
        x=sparse(10*pl(rand(n,1)));
        u=spdiags((sign(pl(rand(m,1)-rand(m,1)))),0,m,m)*(rand(m,1)-rand(m,1));
        b=A*x;
        c=A'
    end
    f = c';
    A = full(A);
    b = full(b);
    lpp = [A eye(m) b;f zeros(1,m+1)];
end
\end{lstlisting}

The program of algorithm \ref{code2} generates a random matrix {\tt A} and the vector {\tt b} for a given number of constraints {\tt m}, number of variables {\tt n} and degree of matrix density {\tt d}. The parameter {\tt d} can vary between 0 to 1, i.e. sparse to very dense. We generated a very dense matrix (0.9), i.e. almost all elements are not zero. The elements of {\tt A} are uniformly distributed between 0 and 50.  The dimension of each problem was defined as a combination of the values: 256, 512, 1024, 2048, 4096 and 8192. With these 6 possible dimensions for the number of variables and number of constraints, we generated and named accordingly 3 different problem instances for each combination, thereby yielding 108 ($6\times 6\times 3$) problems. For example, problem named $512 \times 256$ has 256 variables and 512 constraints with 3 different instances: $512\times 256\_1$, $512 \times256\_2$, $512\times 256\_3$. All problems were generated assuming less-than inequality constraints for maximization.  Therefore, slack variables were always added to the initial tableau such that the actual number of variables were equal to the initial variables plus the number of constraints. So, we always have  more variables than restrictions. Throughout this section, however, we named the problems by the amount of constraint and variables of the non-expanded tableau.

All experiments where performed in a shared-memory server with 4 AMD Opteron 6376 processors with 16 cores, totalizing 64 cores, each running at 2.3GHz, 256 GB DDR3 RAM, 768KB L1 and 16MB L2 individual caches per core, and 16Mbytes L3 caches per socket, running Ubuntu 16.04.2 LTS. The source codes were compiled using GNU GCC 6.3.0. The experiments involved the proposed parallel algorithm with 2, 4, 8, 16, 32 and 64 threads, and our serial implementation of the standard simplex described in Section~\ref{simplex}. We bound the threads to specific cores, using the system variable \texttt{GOMP\_CPU\_AFFINITY}.
%

Table~\ref{values} has the median execution time in seconds over 10 runs of the 3 problems of the same size for all problems and methods analyzed. The first column includes the dimension of the problem, the remaining columns hold the execution times of the proposed parallel simplex implementation with 2, 4, 8, 16, 32 and 64 threads. 
\begin{table}[H]
	\caption{Time in seconds of proposed implementation.}		
	\label{values}
	\centering
	\begin{tabular}{lr|r|r|r|r|r}
		\cline{2-7}
		& \multicolumn{6}{c}{\textbf{Number of threads}} \\ \hline
		\multicolumn{1}{c}{\textbf{Problem}} & \multicolumn{1}{c|}{\textbf{2}} & \multicolumn{1}{c|}{\textbf{4}} & \multicolumn{1}{c|}{\textbf{8}} & \multicolumn{1}{c|}{\textbf{16}} & \multicolumn{1}{c|}{\textbf{32}} & \multicolumn{1}{c}{\textbf{64}} \\ \hline
		\multicolumn{1}{l}{256x256} & 0.036 & 0.027 & 0.021 & 0.031 & 0.092 & 0.301 \\ \hline
		\multicolumn{1}{l}{256x512} & 0.056 & 0.038 & 0.028 & 0.031 & 0.091 & 0.338 \\ \hline
		\multicolumn{1}{l}{256x1024} & 0.086 & 0.052 & 0.038 & 0.038 & 0.082 & 0.286 \\ \hline
		\multicolumn{1}{l}{256x2048} & 0.358 & 0.169 & 0.120 & 0.098 & 0.156 & 0.473 \\ \hline
		\multicolumn{1}{l}{256x4096} & 0.438 & 0.221 & 0.174 & 0.089 & 0.119 & 0.356 \\ \hline
		\multicolumn{1}{l}{256x8192} & 1.351 & 0.627 & 0.547 & 0.347 & 0.291 & 0.532 \\ \hline
		\multicolumn{1}{l}{512x256} & 0.021 & 0.014 & 0.010 & 0.009 & 0.020 & 0.064 \\ \hline
		\multicolumn{1}{l}{512x512} & 0.084 & 0.046 & 0.031 & 0.026 & 0.045 & 0.168 \\ \hline
		\multicolumn{1}{l}{512x1024} & 0.775 & 0.451 & 0.348 & 0.231 & 0.349 & 0.863 \\ \hline
		\multicolumn{1}{l}{512x2048} & 1.155 & 0.662 & 0.589 & 0.332 & 0.342 & 0.891 \\ \hline
		\multicolumn{1}{l}{512x4096} & 3.103 & 1.277 & 1.057 & 0.674 & 0.530 & 0.967 \\ \hline
		\multicolumn{1}{l}{512x8192} & 5.759 & 2.838 & 2.275 & 1.172 & 0.972 & 0.878 \\ \hline
		\multicolumn{1}{l}{1024x256} & 0.327 & 0.168 & 0.131 & 0.060 & 0.071 & 0.236 \\ \hline
		\multicolumn{1}{l}{1024x512} & 1.684 & 1.048 & 0.856 & 0.468 & 0.490 & 1.113 \\ \hline
		\multicolumn{1}{l}{1024x1024} & 2.163 & 1.019 & 0.882 & 0.508 & 0.463 & 0.865 \\ \hline
		\multicolumn{1}{l}{1024x2048} & 1.227 & 0.519 & 0.467 & 0.277 & 0.173 & 0.327 \\ \hline
		\multicolumn{1}{l}{1024x4096} & 38.935 & 20.207 & 15.588 & 6.817 & 3.944 & 4.347 \\ \hline
		\multicolumn{1}{l}{1024x8192} & 90.958 & 63.478 & 40.201 & 45.997 & 46.334 & 12.040 \\ \hline
		\multicolumn{1}{l}{2048x256} & 8.171 & 4.505 & 3.661 & 1.511 & 0.937 & 1.269 \\ \hline
		\multicolumn{1}{l}{2048x512} & 0.799 & 0.541 & 0.421 & 0.244 & 0.114 & 0.128 \\ \hline
		\multicolumn{1}{l}{2048x1024} & 15.935 & 8.810 & 8.703 & 7.249 & 2.680 & 1.742 \\ \hline
		\multicolumn{1}{l}{2048x2048} & 59.883 & 38.809 & 28.364 & 31.859 & 26.620 & 6.708 \\ \hline
		\multicolumn{1}{l}{2048x4096} & 39.007 & 24.344 & 20.444 & 19.423 & 22.391 & 6.058 \\ \hline
		\multicolumn{1}{l}{2048x8192} & 911.651 & 519.929 & 482.603 & 405.369 & 332.439 & 268.802 \\ \hline
		\multicolumn{1}{l}{4096x256} & 54.202 & 33.635 & 24.085 & 24.954 & 25.742 & 17.898 \\ \hline
		\multicolumn{1}{l}{4096x512} & 46.711 & 31.147 & 23.529 & 22.518 & 23.134 & 19.016 \\ \hline
		\multicolumn{1}{l}{4096x1024} & 197.108 & 133.184 & 92.854 & 79.553 & 91.360 & 67.271 \\ \hline
		\multicolumn{1}{l}{4096x2048} & 0.528 & 0.435 & 0.377 & 0.343 & 0.323 & 0.305 \\ \hline
		\multicolumn{1}{l}{4096x4096} & 1382.832 & 888.342 & 799.383 & 592.892 & 622.866 & 373.573 \\ \hline
		\multicolumn{1}{l}{4096x8192} & 1071.709 & 671.992 & 528.183 & 398.582 & 442.300 & 266.595 \\ \hline
		\multicolumn{1}{l}{8192x256} & 274.244 & 150.246 & 119.035 & 104.966 & 104.110 & 65.755 \\ \hline
		\multicolumn{1}{l}{8192x512} & 18.183 & 11.740 & 10.290 & 9.970 & 10.635 & 9.000 \\ \hline
		\multicolumn{1}{l}{8192x1024} & 126.048 & 72.757 & 60.564 & 57.594 & 53.484 & 36.189 \\ \hline
		\multicolumn{1}{l}{8192x2048} & 52.146 & 28.557 & 26.544 & 24.463 & 25.285 & 16.443 \\ \hline
		\multicolumn{1}{l}{8192x4096} & 944.021 & 544.355 & 485.721 & 385.115 & 344.792 & 201.071 \\ \hline
		\multicolumn{1}{l}{8192x8192} & 25208.584 & 16344.648 & 13628.544 & 11215.784 & 9979.548 & 5458.975 \\ \hline
	\end{tabular}
\end{table}

\subsection{Speedup}
\label{sec:speedup}
%
%
%
%
%

In this Section, we discuss the speedup performance of the proposed algorithm. The speedup is a measure of how much faster the parallel code runs compared to the sequential code. Due to the excessive amount of time that it was necessary to run the sequential simplex algorithm, for all speedup and parallel efficiency calculations, we have assumed that the sequential execution time was twice the execution time of the case with 2 threads. By that, we are assuming a linear speedup for 2 threads and embracing in our results any potential performance inaccuracy arising from that.

\begin{figure*}[!ht]
	\centering
	\subfloat[Problems with 256 constraints.]{\includegraphics[width=.49\textwidth]{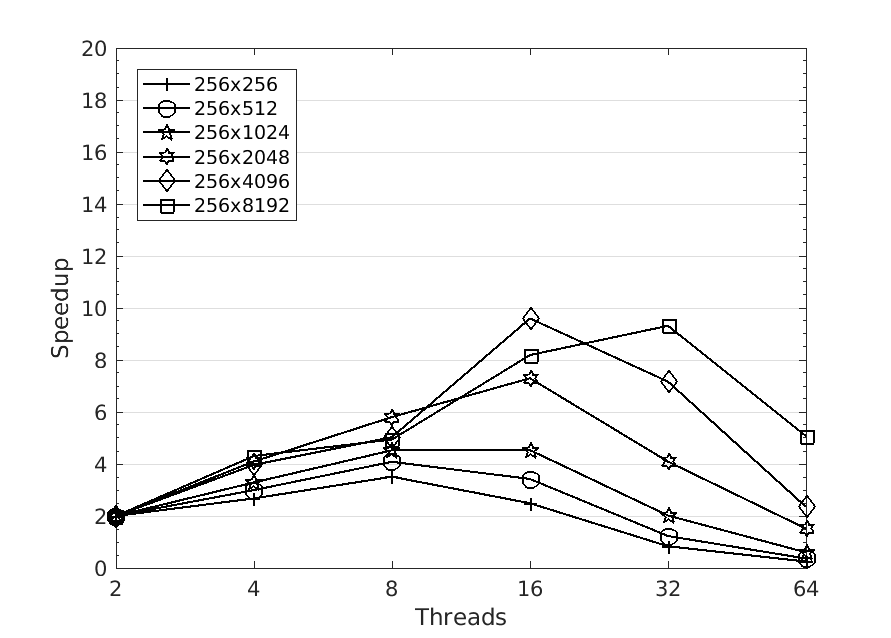}
		\label{fig:speedup256c}}
	\hfil
	\subfloat[Problems with 256 variables.]{\includegraphics[width=.49\textwidth]{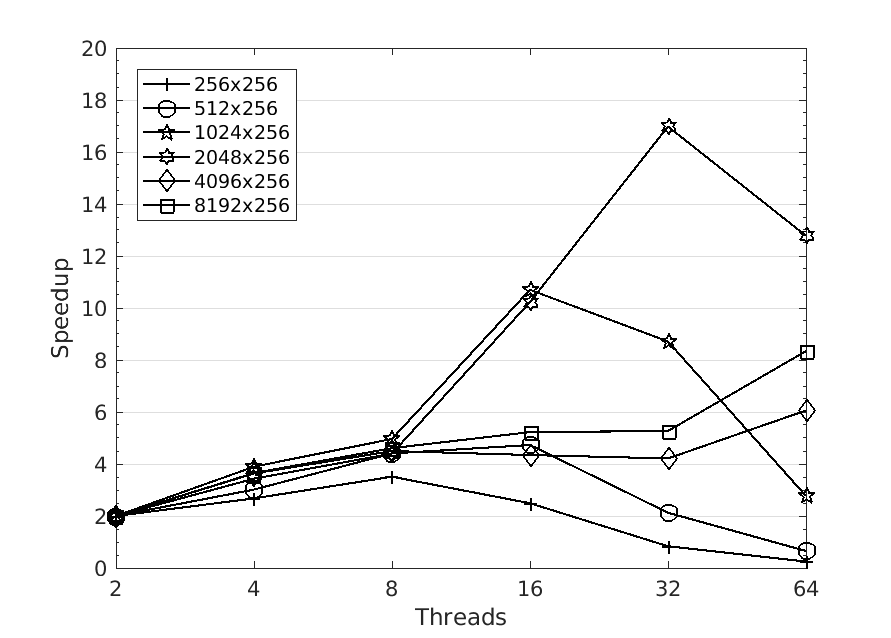}
		\label{fig:speedup256v}}

	\caption{Speedups of the proposed implementation for fixing one of the problem dimensions in 256 for primal (a) and dual (b) formulations.}
	\label{fig:speedup2}	
\end{figure*}

\figurename~\ref{fig:speedup256c} shows speedups for when we fix the constraints at 256. Notice that the speedup starts equal to two, because of our assumption that the serial time is twice the time for two threads. Observe in the figure that, as the speedup increases and then drops for each problem, it decrease at larger amounts of threads for larger problems. This is an indication that the performance loss is due to the overhead of adding more threads, which does not pay off for smaller problems. In addition, from 32 threads on, the largest speedups were those with the largest amount of variables. This may indicate that the parallel portion of the code is increasing with amount of variables. \figurename~\ref{fig:speedup256v} shows speedup for when variables are fixed at 256. The same pattern of \figurename~\ref{fig:speedup256c} can be observed. Speedup drops happen at larger number of threads though.

\begin{figure*}[!ht]	
	\subfloat[Problems with 512 constraints.]{\includegraphics[width=.49\textwidth]{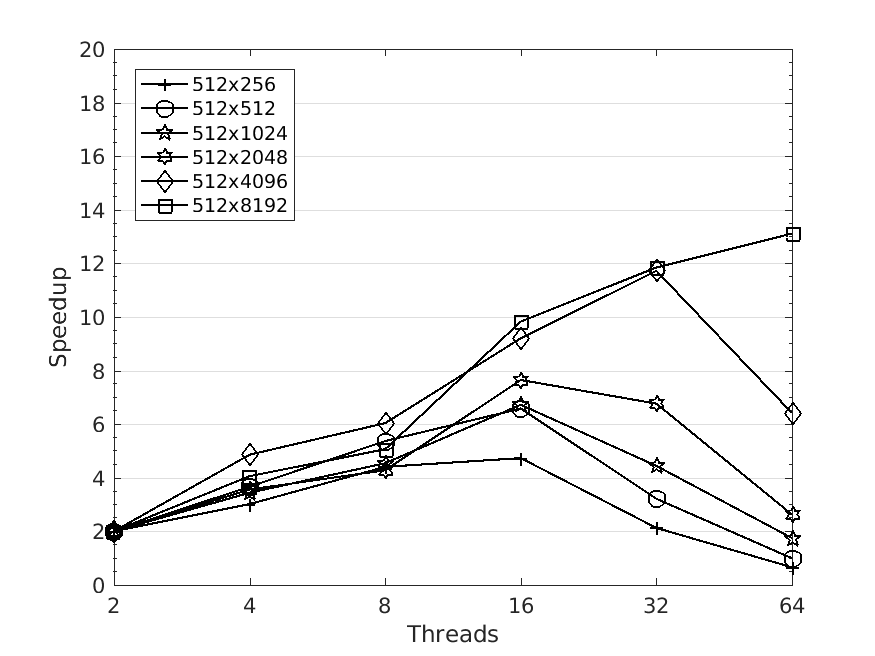}
		\label{fig:speedup512c}}
	\hfil	
	\subfloat[Problems with 512 variables.]{\includegraphics[width=.49\textwidth]{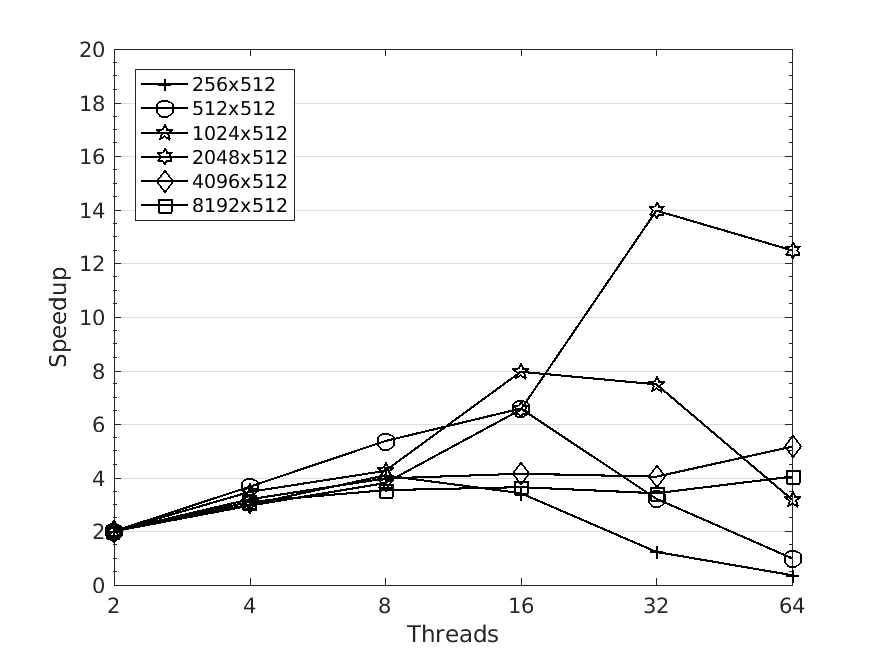}
		\label{fig:speedup512v}}	
	
	\caption{Speedups of the proposed implementation for fixing one of the problem dimensions in 512 for primal (a) and dual (b) formulations.}	
	\label{fig:speedup3}
\end{figure*}

\figurename~\ref{fig:speedup512c} shows speedups fixing the number of constraints at 512. All problems with less than 4096 variables have a drop at 32 threads. The drop for the second largest happens at 64 threads, while it was not observed for the largest problem. At 64 treads, we can see that the larger amounts of variables the larger the speedup, indicating that increasing the number of variables improves speedup performance. Similarly, the same patterns of \figurename~\ref{fig:speedup256v} appear in \figurename~\ref{fig:speedup512v}, where the number of variables is fixed at 512. Here, too, the drops of speedup happens at larger number of threads. Notice that formulating the problems this way, all problems, but the $256\times256$, have the number of variables increased by the number of constraints to accommodate the slack variables. This causes the problems to grow faster in size when the number of constraints increase than in the other formulation when the number of variables increase. For this reason, the speedup drop caused by the extra threads happens at larger amounts of threads in this formulation. Except for the two largest problems, which do not experience the same speedup increase as the two smaller problems ($1024\times256$ and $2048\times256$) neither the same speedup drop as all other problems. For being much larger that the other problems, the speedup of these problems are probably limited by the insufficient amount of cache memory. This also explains their expressive speedup increase when going from 32 to 64 threads, because of the increase in the number of L3 caches used in the computation that goes from 2 to 4.

\begin{figure*}[!ht]	
	\subfloat[[Problems with 1024 constraints.]{\includegraphics[width=.49\textwidth]{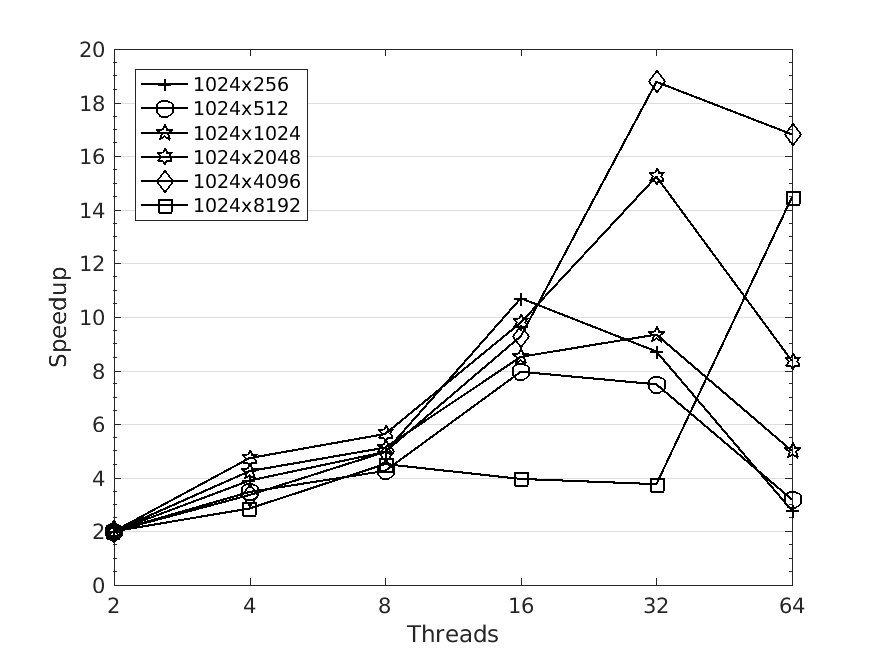}
		\label{fig:speedup1024c}}
	\hfil	
	\subfloat[Problems with 1024 variables.]{\includegraphics[width=.49\textwidth]{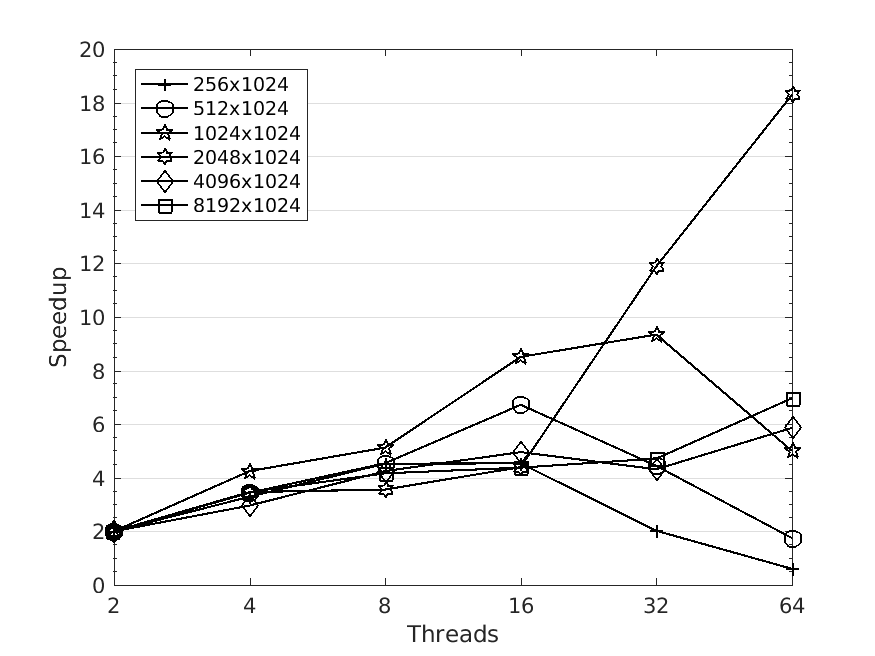}
		\label{fig:speedup1024v}}	
	\caption{Speedups of the proposed implementation for fixing one of the problem dimensions in 1024 for primal (a) and dual (b) formulations.}
	\label{fig:speedup4}	
\end{figure*}

\begin{figure*}[!ht]
	\centering
	\subfloat[Problems with 2048 constraints.]{\includegraphics[width=.49\textwidth]{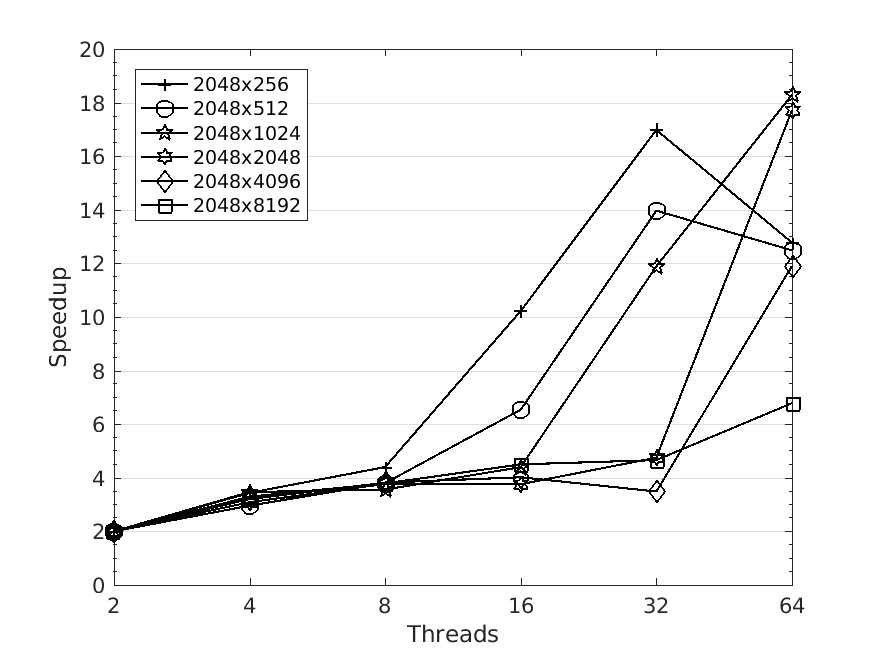}
		\label{fig:speedup2048c}}
	\hfil
	\subfloat[Problems with 2048 variables.]{\includegraphics[width=.49\textwidth]{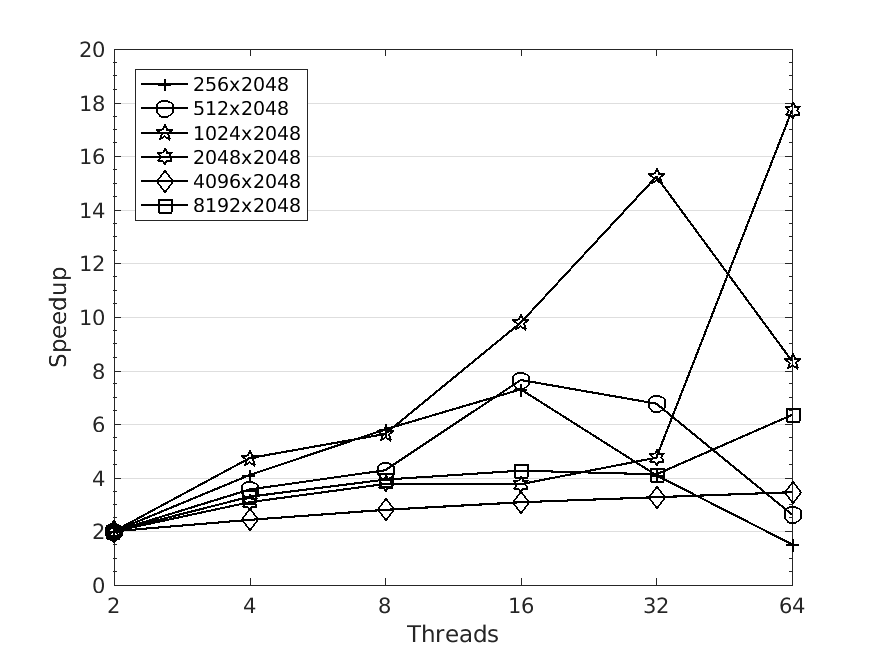}
		\label{fig:speedup2048v}}
	\caption{Speedups of the proposed implementation for fixing one of the problem dimensions in 2048 for primal (a) and dual (b) formulations.}
	\label{fig:speedup5}
\end{figure*}

\begin{figure*}[!ht]	
	\subfloat[Problems with 4096 constraints.]{\includegraphics[width=.49\textwidth]{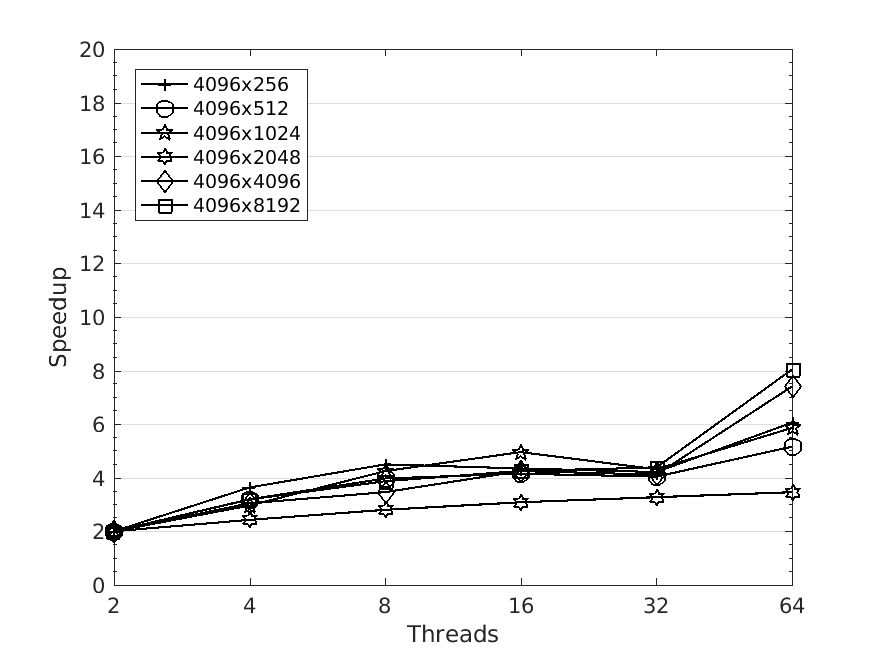}
		\label{fig:speedup4096c}}
	\hfil	
	\subfloat[Problems with 4096 variables.]{\includegraphics[width=.49\textwidth]{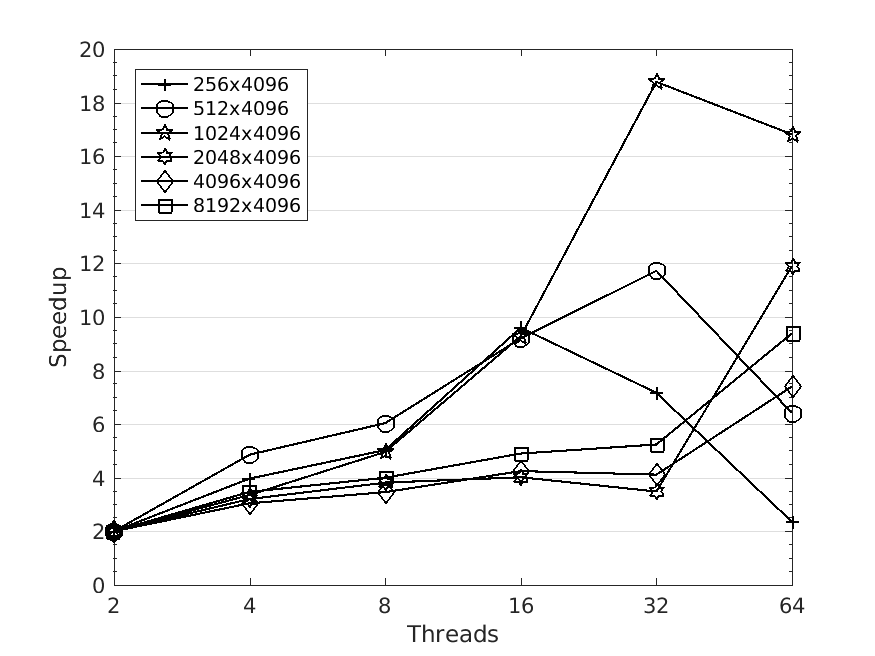}
		\label{fig:speedup4096v}}	
	
	\caption{Speedups of the proposed implementation for fixing one of the problem dimensions in 4096 for primal (a) and dual (b) formulations.}
	\label{fig:speedup6}
\end{figure*}

\begin{figure*}[!ht]
	\centering
	
	\subfloat[Problems with 8192 constraints.]{\includegraphics[width=.49\textwidth]{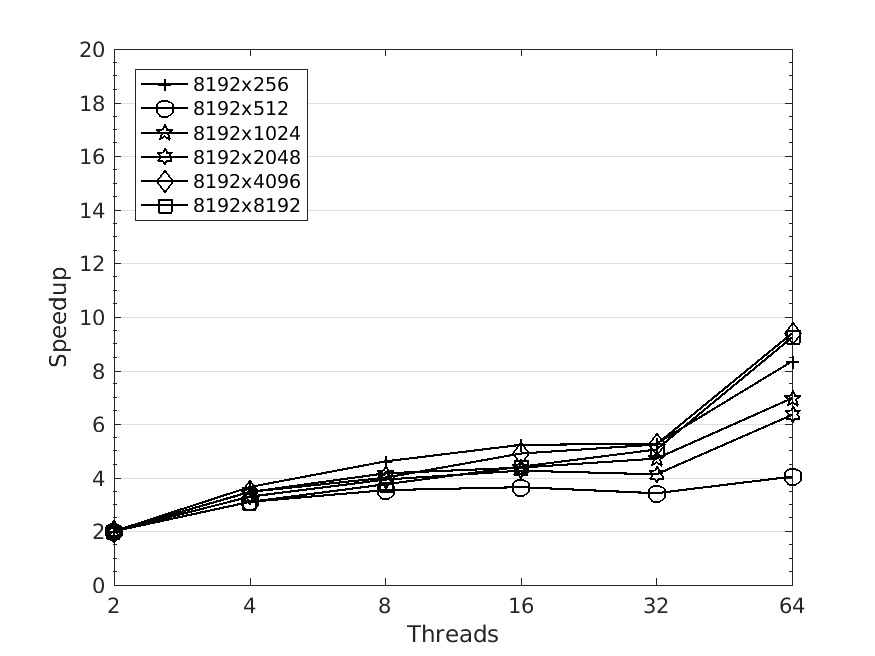}
		\label{fig:speedup8192c}}
	\hfil	
	\subfloat[Problems with 8192 variables.]{\includegraphics[width=.49\textwidth]{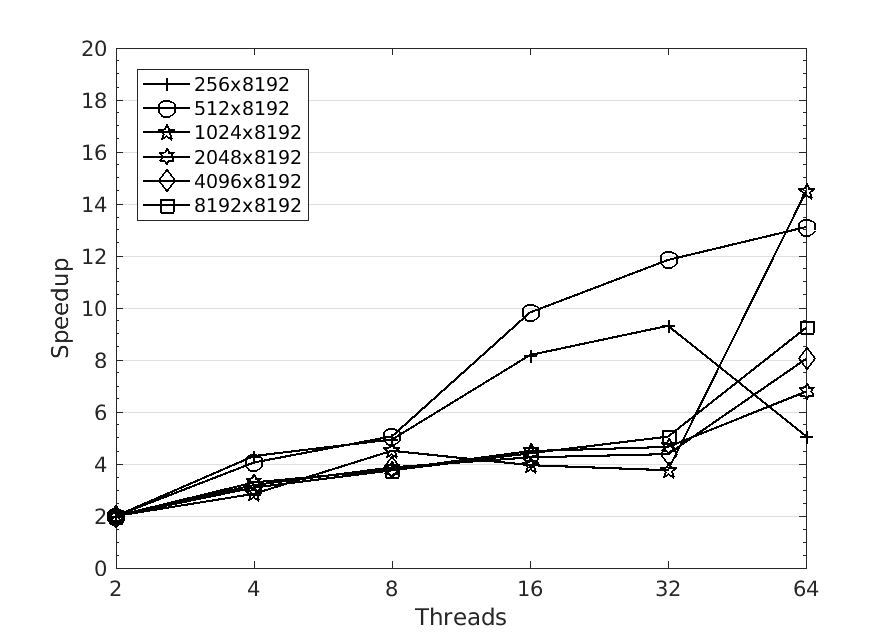}
		\label{fig:speedup8192v}}	
	
	\caption{Speedups of the proposed implementation for fixing one of the problem dimensions in 8192 for primal (a) and dual (b) formulations.}
	\label{fig:speedup7}
\end{figure*}

The patterns of Figs. \ref{fig:speedup2} and \ref{fig:speedup3} repeat throughout Figs. \ref{fig:speedup4} to \ref{fig:speedup7}. These patterns are characterized by a moderate increase in speedup followed by a drop in the number of threads gets larger. Most likely, the speedup drops have to cause. First, for smaller problems, increasing the number of threads causes the threading overhead to become more significative than the gains in concurrency. For larger problems, however, the drop cannot be attributed to the same cause because the workload assigned to each thread dilutes the extra threads overhead. So, the second cause can be attributed to insufficient cache memory to store the simplex table. The largest evidence for that in some plots of Figs. 2 to 7 is the sudden speedup rise when going from 32 to 64 threads, when the amount of L3 caches goes from 2 to 4. Note that this rise in the speedups happens even when the drop due to threading overhead was already observed for smaller amounts of threads.

\subsection{Parallel Efficiency and the limits of cache}
\begin{figure*}[!ht]
	\centering
	\subfloat[Problems with 256 constraints.]{\includegraphics[width=.49\textwidth]{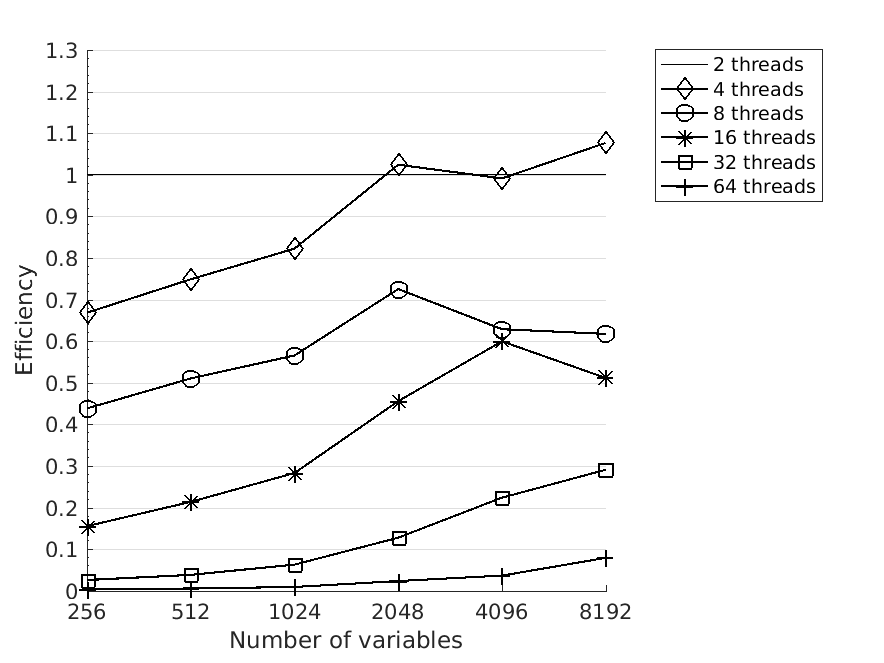}
		\label{fig:eficiency256c}}
	\hfil
	\subfloat[Problems with 256 variables]{\includegraphics[width=.49\textwidth]{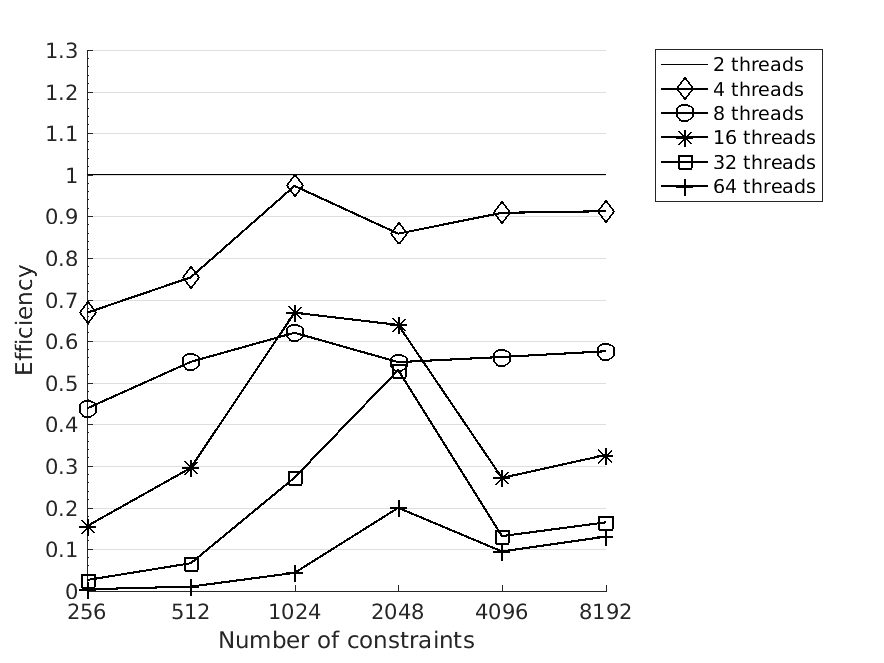}
		\label{fig:eficiency256v}}
	
	\caption{Parallel efficiency of the proposed implementation for fixing one of the problem dimensions in 256 for primal (a) and dual (b) formulations.}
	\label{fig:eficiency8}
	
	\end{figure*}

\begin{figure*}[!ht]
	\subfloat[Problems with 512 constraints.]{\includegraphics[width=.49\textwidth]{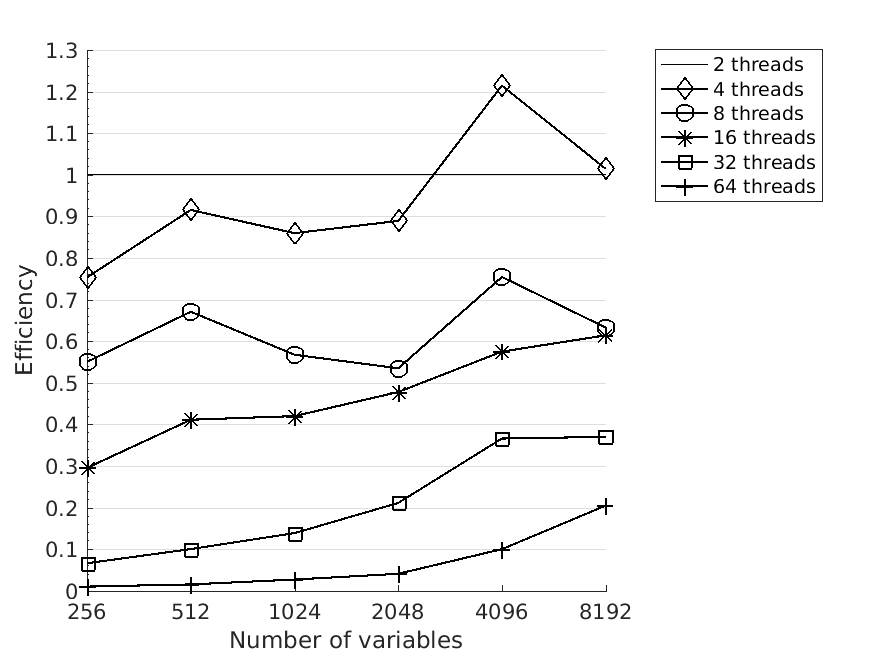}
		\label{fig:eficiency512c}}
	\hfil	
	\subfloat[Problems with 512 variables.]{\includegraphics[width=.49\textwidth]{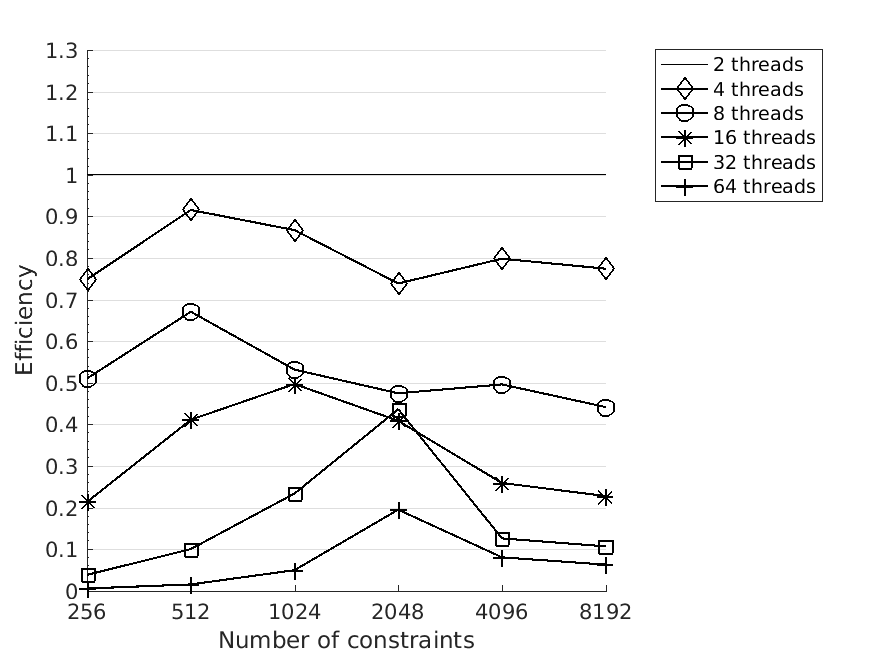}
		\label{fig:eficiency512v}}	
	
	\caption{Parallel efficiency of the proposed implementation for fixing one of the problem dimensions in 512 for primal (a) and dual (b) formulations.}
    \label{fig:eficiency9}	
	\end{figure*}

\begin{figure*}[!ht]
	\subfloat[Problems with 1024 constraints.]{\includegraphics[width=.49\textwidth]{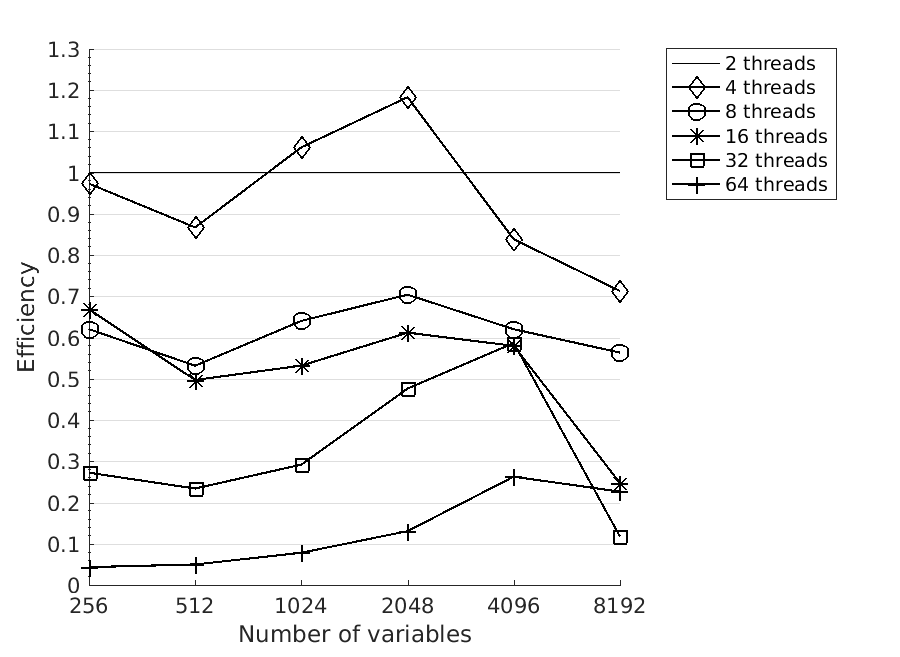}
		\label{fig:eficiency1024c}}
	\hfil	
	\subfloat[Problems with 1024 variables.]{\includegraphics[width=.49\textwidth]{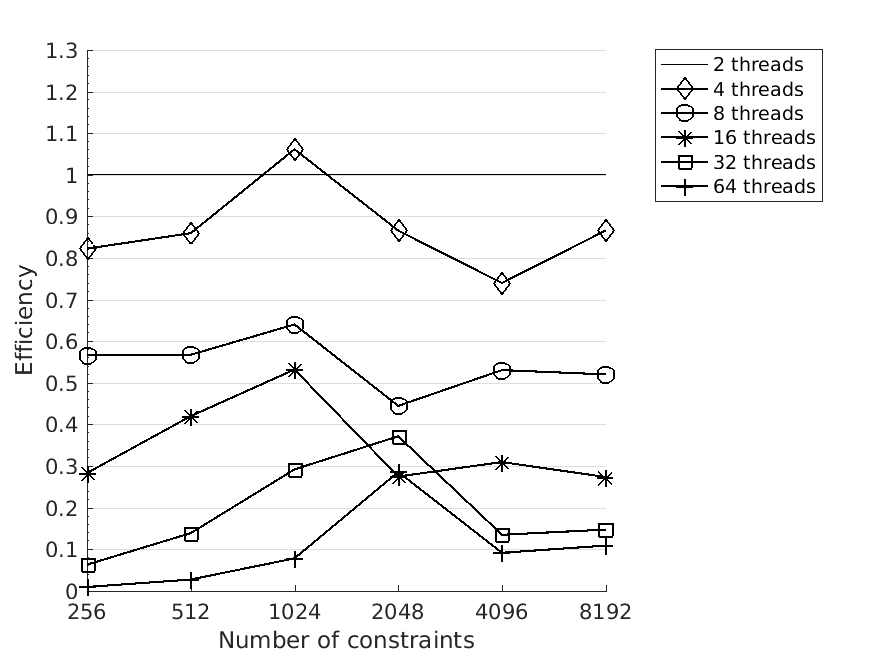}
		\label{fig:eficiency1024v}}	
	
	\caption{Parallel efficiency of the proposed implementation for fixing one of the problem dimensions in 1024 for primal (a) and dual (b) formulations.}
	\label{fig:eficiency10}	
\end{figure*}

\begin{figure*}[!ht]
	\centering
	\subfloat[Problems with 2048 constraints.]{\includegraphics[width=.49\textwidth]{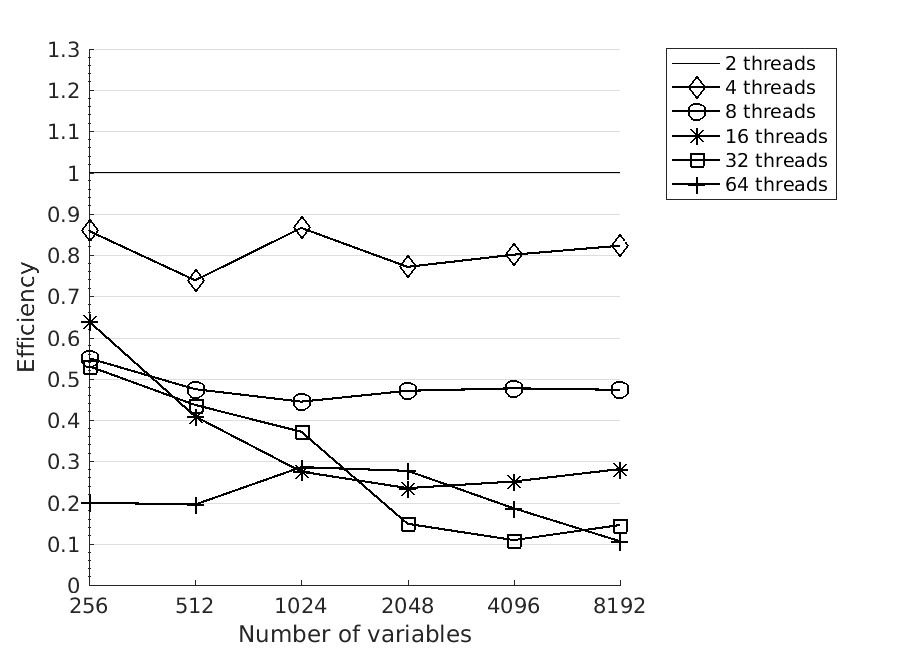}
		\label{fig:eficiency2048c}}
	\hfil
	\subfloat[Problems with 2048 variables.]{\includegraphics[width=.49\textwidth]{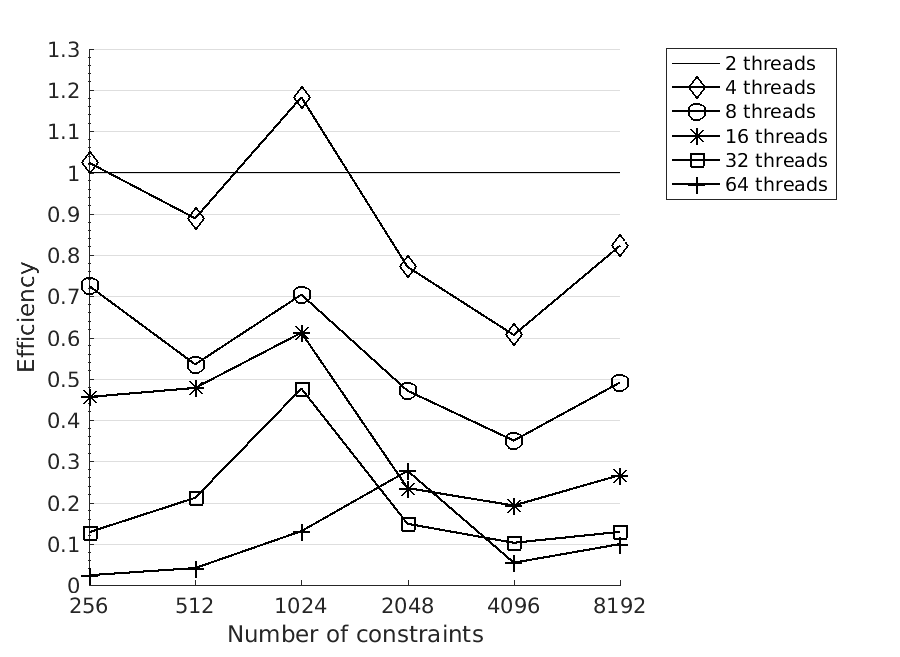}
		\label{fig:eficiency2048v}}

	\caption{Parallel efficiency of the proposed implementation for fixing one of the problem dimensions in 2048 for primal (a) and dual (b) formulations.}
	\label{fig:eficiency11}	
	\end{figure*}

\begin{figure*}[!ht]
	\subfloat[Problems with 4096 constraints.]{\includegraphics[width=.49\textwidth]{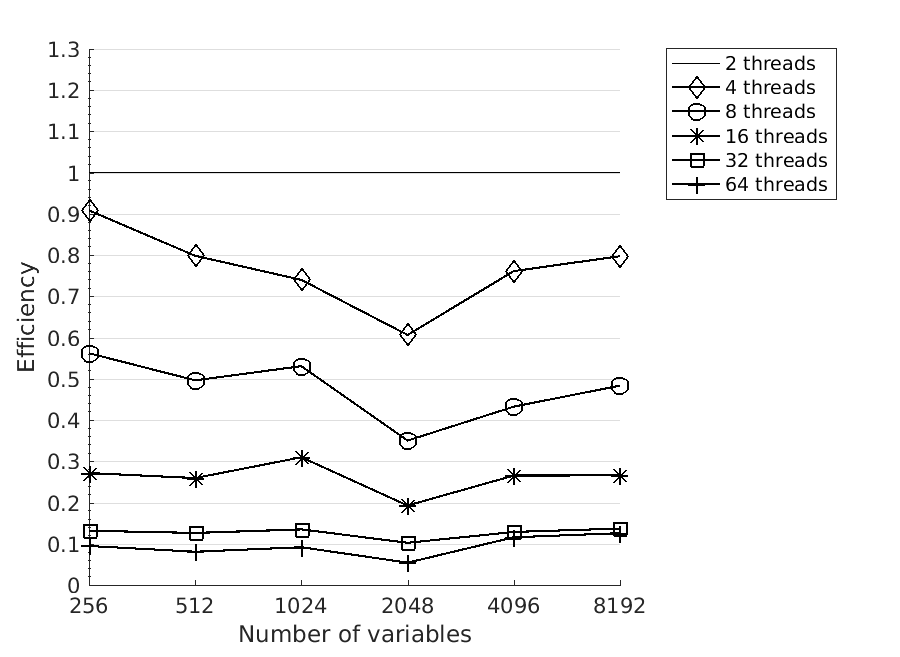}
		\label{fig:eficiency4096c}}
	\hfil	
	\subfloat[Problems with 4096 variables.]{\includegraphics[width=.49\textwidth]{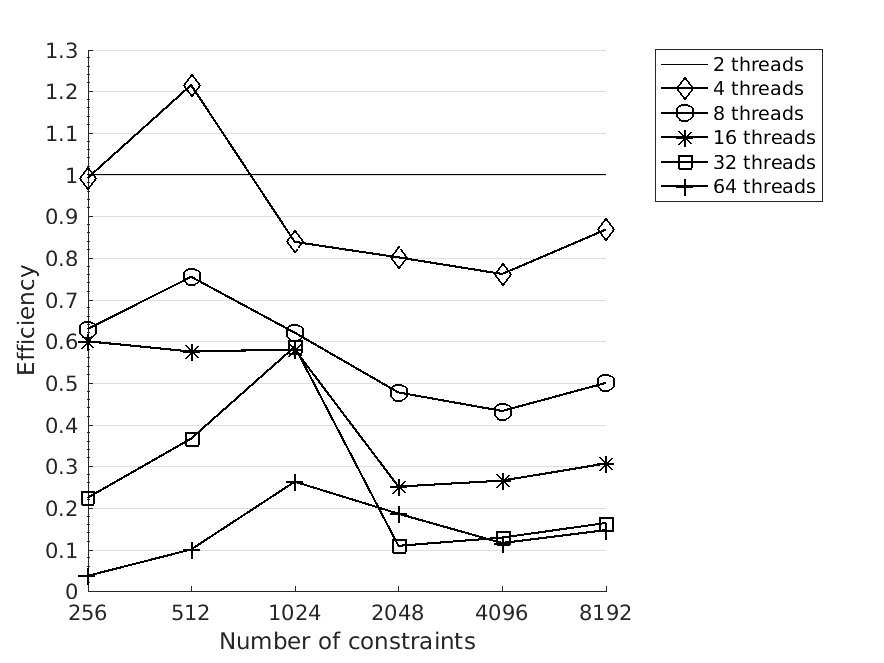}
		\label{fig:eficiency4096v}}	
	
	\caption{Parallel efficiency of the proposed implementation for fixing one of the problem dimensions in 4096 for primal (a) and dual (b) formulations.}
	\label{fig:eficiency12}	
	
	\end{figure*}

\begin{figure*}[!ht]
	\subfloat[Problems with 8192 constraints.]{\includegraphics[width=.49\textwidth]{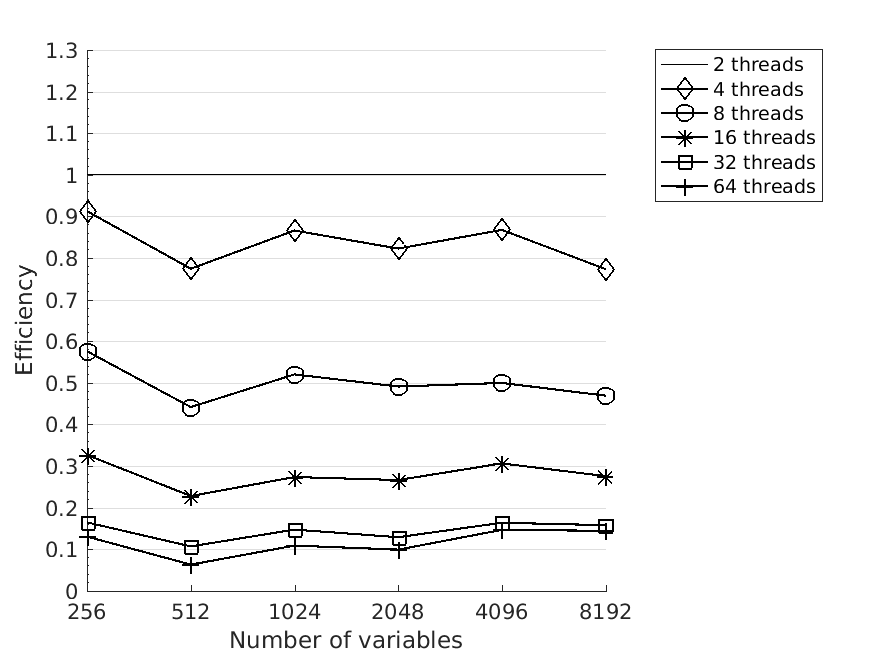}
		\label{fig:eficiency8192c}}
	\hfil	
	\subfloat[Problems with 8192 variables.]{\includegraphics[width=.49\textwidth]{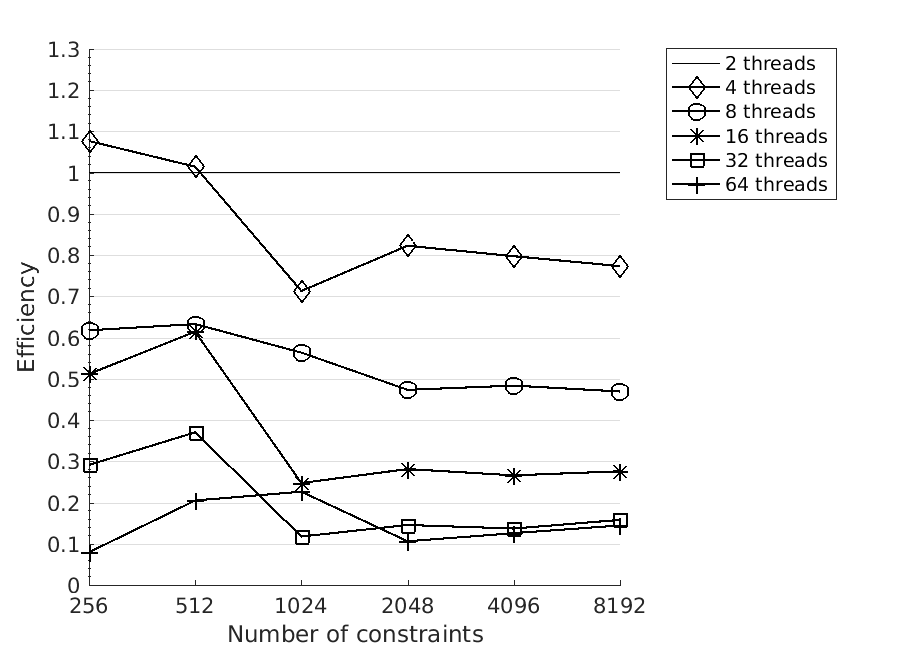}
		\label{fig:eficiency8192v}}	
	
	\caption{Parallel efficiency of the proposed implementation for fixing one of the problem dimensions in 8192 for primal (a) and dual (b) formulations.}
	\label{fig:eficiency13}	
\end{figure*}

In this subsection, we present the plots for parallel efficiency and compute the limits of problem sizes that make efficient use of cache. Parallel efficiency normalized measure for speedup. It indicates how effectively each processor element is used. Figs. \ref{fig:eficiency8}-\ref{fig:eficiency13} present the parallel efficiencies of all problems and their primal/dual counterpart. In these normalized versions of speedups, the effect of cache-size limits is even more clear than in Figs.\ref{fig:speedup2}-\ref{fig:speedup7}. Throughout the figures, increasing the dimensions of the problem causes an efficiency rise followed by a drop. This effect becomes more predominant at lower number of variables and constraints as we move along from Fig. \ref{fig:eficiency8} to Fig.\ref{fig:eficiency13}.

\begin{figure*}[!ht]
	\centering	 
	\includegraphics[width=\textwidth]{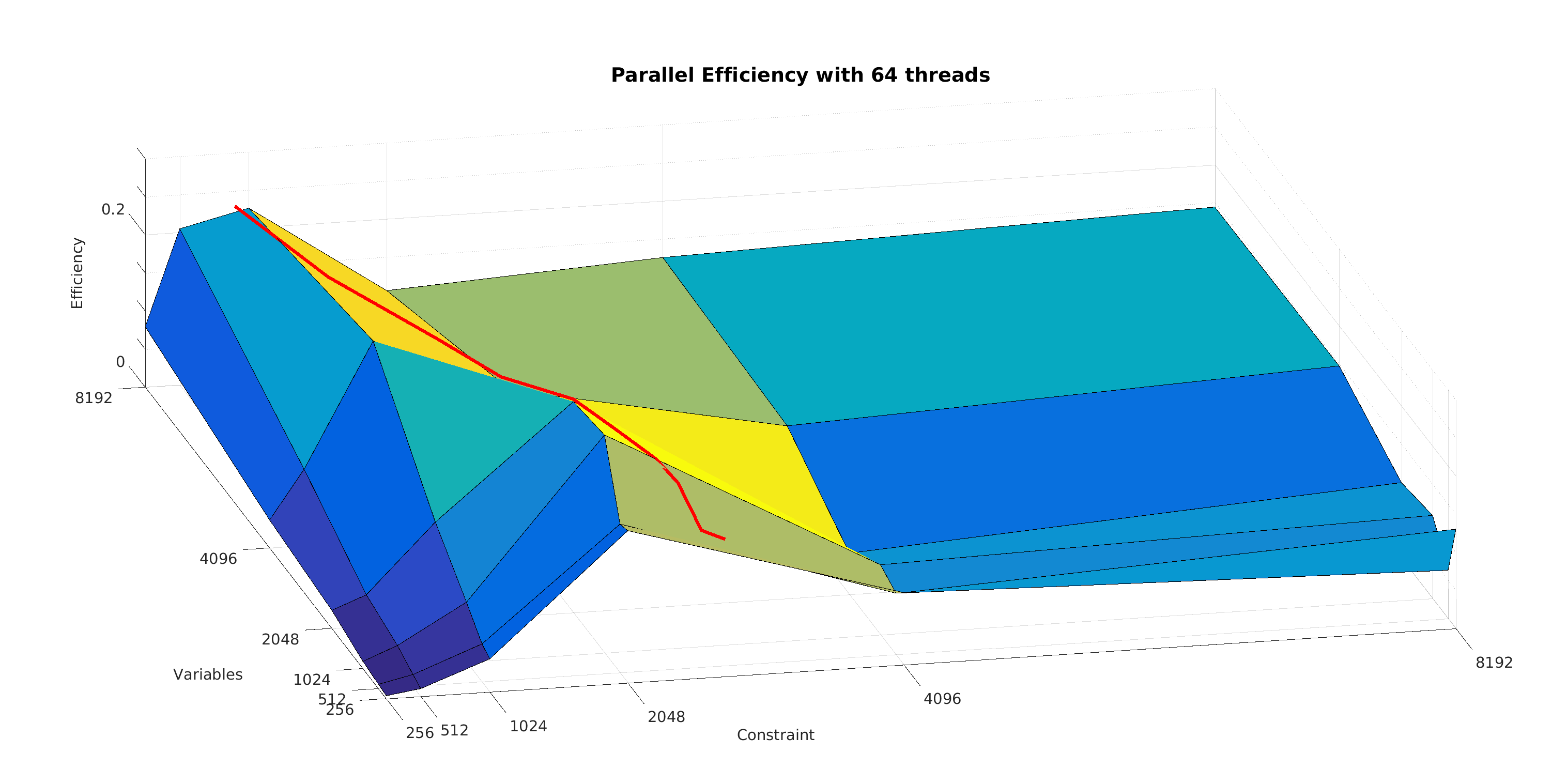}
	\caption{Parallel Efficiency for 64 threads. The red line delimits the problem sizes that reach approximately the full use of all L3 caches.} 
	\label{tablecache}
\end{figure*}

In Fig. \ref{tablecache}, the effect of cache-size limits can be clearly observed. 
The figure draws the efficiency of the proposed parallel algorithm with 64 threads for varying constraints and variables. 

As problem sizes increases, we observe the efficiency scaling up and then it drops regardless of weather the increase is due to more variables or more constraints. There is not enough cache space to accommodate  larger problems, causing data to be fetched more frequently from the main memory. This limit can be estimated by calculating the amount of variables and constraints that fits in the four L3 caches. The total amount of bytes being used by the LP problem cannot exceed 64 Mbytes, which is the total size of the four L3 caches. Considering a double floating point precision where each value requires 8 bytes for storage, it follows that:
	\begin{align}
		(n + b) \times n \times 8{\rm bytes} &\leq 64 {\rm Mbytes},
	\end{align}
where $n$ is the amount of constraints and $b$ the amount of variables. Solving for $n$ we find the maximum number of constraints relative to a given number of variables $b$ that avoids cache resource limitation, as follows.
\begin{align}
		n^{2} + bn  &\leq \frac{64{\rm Mbytes}}{8{\rm bytes}}, \nonumber\\
		n^{2} + bn &\leq 8388608, \nonumber\\
		n^{2} + bn - 8388608 &\leq 0, \nonumber\\
		n &\leq \frac{-b\pm \sqrt{b^2 + 33554432}}{2}.
    	\label{eqcache}
	\end{align}
Replacing $b$ by the values of the variables used in the problems dimensions, i.e. 256, 512, 1024, 2048, 4096 and 8192, (\ref{eqcache}) gives as maximum numbers of constraints 2771, 2651, 2429, 2048, 1499 and 1499, respectively. For example, for 1024 constraints, the number of variables cannot exceed 2429. The red line (Fig. \ref{tablecache}) shows the function described by (2), which roughly estimates the limit of the amount of variables
and constraints that fits in the L3 caches.

Noticeably, in Fig. \ref{tablecache} indicating that the algorithm is scalable up to a specific problem size. Scalable problems sustain efficiency as the problem size increases. Past a limit, the proposed parallel implementation fails to sustain efficiency for larger problems due to cache resource limitation.


%

%% file: tex/consideration.tex
\section{Conclusions}

This paper presents a parallel implementation of the standard Simplex algorithm for multicore processors for solving large-scale LP problems. We described the general scheme of the parallelization, explaining each step of the standard simplex algorithm and detailing important optimization steps of our implementation. We compared the speedups of the proposed parallel algorithm against the standard simplex implementation, using up to 64 threads in a shared memory machine with 64 cores. The proposed parallel simplex algorithm demonstrated scalable performance for various combinations of variables and constraints. We have shown that the implementation is weakly scalable up to the limits of efficient use of caches. We have divided the equations that describe this limit and verified it experimentally.

%
%


Most probably this limit to the scalability can be improved by using techniques to maximize cache locality such as a blocking and red polyhedral transformations, which we believe is worthwhile investigating in the future.  